\documentclass{aa}
\usepackage{psfig}

\def\bron{SAX~J2103.5+4545}
\def\ecs{erg~cm$^{-2}$s$^{-1}$}

\begin{document}
\thesaurus{01(08.02.1; 08.09.2 HD~200709; 08.14.1; 08.16.7 \bron; 13.25.5)}

\title{Discovery of the transient X-ray pulsar \bron}
\author{F. Hulleman, J.J.M.~in~'t Zand \and J. Heise}
\offprints{J.J.M.~in~'t Zand}

\institute{   Space Research Organization Netherlands, Sorbonnelaan 2,
              3584 CA Utrecht, the Netherlands
                        }
\date{Received, accepted }
\titlerunning{\bron}
\maketitle

\begin{abstract}
We report the discovery of the X-ray transient \bron\ which
was active from February to September 1997. The observed peak intensity of 20 mCrab (2 to 25~keV) occurred on April 11. 
An analysis of data obtained around the time of the peak revealed a pulsed signal with a period of $358.61\pm0.03$~s on MJD~50569. 
The pulse profile has a pulsed fraction of $\sim40$\%. 
No change in the pulse period was detected, with an upper limit of 
$6$~s~yr$^{-1}$. The energy spectrum complies to a power law 
function with a photon index of $1.27\pm0.14$ and low-energy absorption equivalent 
to a hydrogen column density of $(3.1\pm1.4)\times10^{22}$ atoms cm$^{-2}$ of cold 
gas of cosmic abundances. In analogy to other X-ray pulsars with similar 
characteristics we propose this object to be a neutron star in close orbit 
around a mass-losing star of early spectral type. The B star HD~200709 is a 
marginal candidate optical counterpart.

\end{abstract}

\section{Introduction}
\label{intro}

Currently, about 50 accretion-powered X-ray pulsars are known
(e.g., \cite{batse}, \cite{jvp95}). The majority of these are 
located in binary systems (a few cases are thought to accrete 
from molecular clouds, e.g., Corbet et al. 1995). If optically 
identified, the companion 
is usually a high-mass star. In these cases half are giant or 
supergiant stars and half are Oe or Be type stars. Interestingly,
the subset of those that are X-ray transient all have Oe or Be type 
companions. This testifies in favor of the notion that such stars
loose matter in an irregular fashion (e.g., \cite{Heuvel}).

From the typical intrinsic luminosities and distances of
X-ray pulsars, it is expected that a relatively
large population of faint (i.e., fainter than 10$^{-9}$~\ecs) 
X-ray pulsars exists in our galaxy. In fact, due to successful observations
with {\em Ginga}, the Rossi X-ray Timing Explorer and {\em Asca}
this has been confirmed (e.g., \cite{AX J1820}, \cite{scutum}, Marsden 
et al. 1998). Furthermore, Koyama et al. (1990) observed a 
concentration of \mbox{X-ray} pulsars in the Scutum arm, very likely
associated with a region of recent star formation. 

We report the discovery of another faint transient X-ray pulsar using data 
accumulated by the Wide Field Cameras on board the BeppoSAX satellite. We 
present the observations and detection in section~\ref{secobs}, the analysis 
of the pulsar signal in section~\ref{sec:period}, and the energy spectrum in 
section~\ref{secspec}. We discuss the implications of the
details of the measurements for the nature of the source in 
section~\ref{sec:discuss}. 

\section{Observations, detection and time profile}
\label{secobs}

The observations were performed with the Wide Field
Camera (WFC) instrument (Jager et al. 1997). This instrument consists of two
coded aperture cameras that point in opposite directions at the X-ray
sky. They are located on the BeppoSAX X-ray satellite (e.g., Boella et al.
1997) which is in operation since April 1996. Each camera is equipped with
a multi-wire proportional xenon counter with a bandpass of 2 to 25 keV.
The field of view is 40 by 40 square degrees (3.7\% of the sky). 
The angular resolution is 5\arcmin, the accuracy to position a point 
source is 2 to 3 times better than that.

About 90\% of the observations with WFC are carried out in secondary mode.
In this mode, observations with the narrow-field instruments on BeppoSAX
dictate the orientation of the satellite and, thus, the pointing 
direction of the WFC. The Cygnus field is frequently covered
by secondary mode observations. During the first 1.5 yr of the mission
the WFCs were exposed to it during 90 days, with a fairly
uniform time coverage. The total effective exposure time (i.e., when the Earth 
is outside the field of view) is $\sim2\times10^6$~s.

\begin{figure}[t]
\psfig{figure=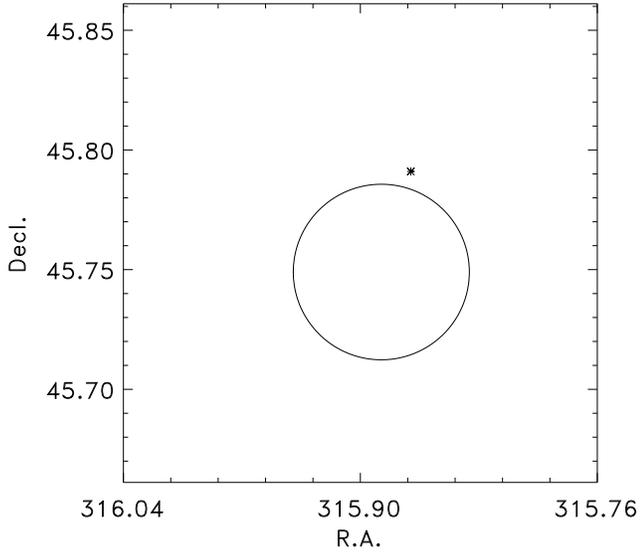,width=\columnwidth,clip=t}

\caption[]{Error box of \bron\ at 99\% confidence level, based on
2 to 10 keV data from two observation periods. Systematic
and statistical sources of error are taken into account, the systematic
errors dominate. The asterix points to HD~200709.
\label{figerrbox}}
\end{figure}

During April 1997, a faint (i.e., for WFC standards) transient was detected 
in secondary mode observations of the Cygnus field, 7.4 degrees from the 
bright X-ray source Cyg X-3. The error box
of this transient is presented in fig.~\ref{figerrbox}. A search of this box
with the Simbad database gave a negative result on an identification with
any cataloged X-ray source within 20\arcmin from the best fit position. We conclude 
that this is a previously unknown X-ray transient. The best fit position is 
R.A.~=~21$^{\rm h}~03^{\rm m}~33^{\rm s}$, 
Decl.~=~+45$^{\rm o}$45\farcm0 (Eq. 2000.0). We designate the source \bron. 
The source is close to the Galactic plane, at $b^{\rm II}=-0.64^{\rm o}$.
A search for catalogued optical counterparts revealed a candidate at 2.6\arcmin
from the best fit position. Although it is just outside the error box,
we regard it initially as a candidate to be considered. It is HD~200709, 
a B8V star with $m_{\rm v}=9.21$ (Bouigue 1959).

\begin{figure}[t]
\psfig{figure=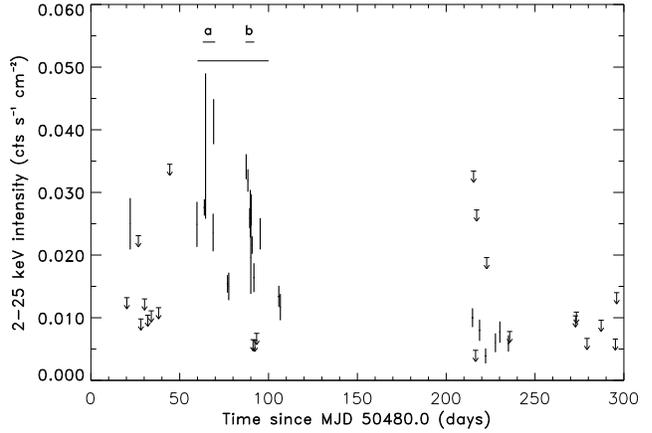,width=\columnwidth,clip=t}

\caption[]{Time profile of \bron. The starting point of the time axis is
February 1st, 1997. The thick vertical lines indicate the 
error bars on intensity for detections, the thin arrows indicate 3$\sigma$ 
upper limits. The time resolution is that of one observation period.
This varies between 5 min and 1.15 days. The changing sensitivity is caused
by changing off-axis angles for the source. 0.01~cts~s$^{-1}$cm$^{-2}$ is 5 
mCrab in 2 to 25 keV. The long horizontal line indicates the times for which
a timing and spectral analysis was performed (see sections \ref{sec:period}
and \ref{secspec}).
\label{figlc}}
\end{figure}

Fig.~\ref{figlc} presents a time history of the intensity of \bron. 
The source was active for almost 8 months, from February to 
September 1997. No other detections occurred in the complete data set from 
June 1996 to March 1998. The observed  peak intensity was 
about 0.04~cts~s$^{-1}$cm$^{-2}$ which equals about 20~mCrab (2 to 25 keV).
There is considerable variability on time scales of hours and
days which is exemplified around MJD~50570 when 3$\sigma$ upper limit follow
detections that are a factor 3 as high. 
No X-ray bursts were detected.

\section{Period determination and pulse shape} \label{sec:period}

\begin{figure}[t]
\psfig{figure=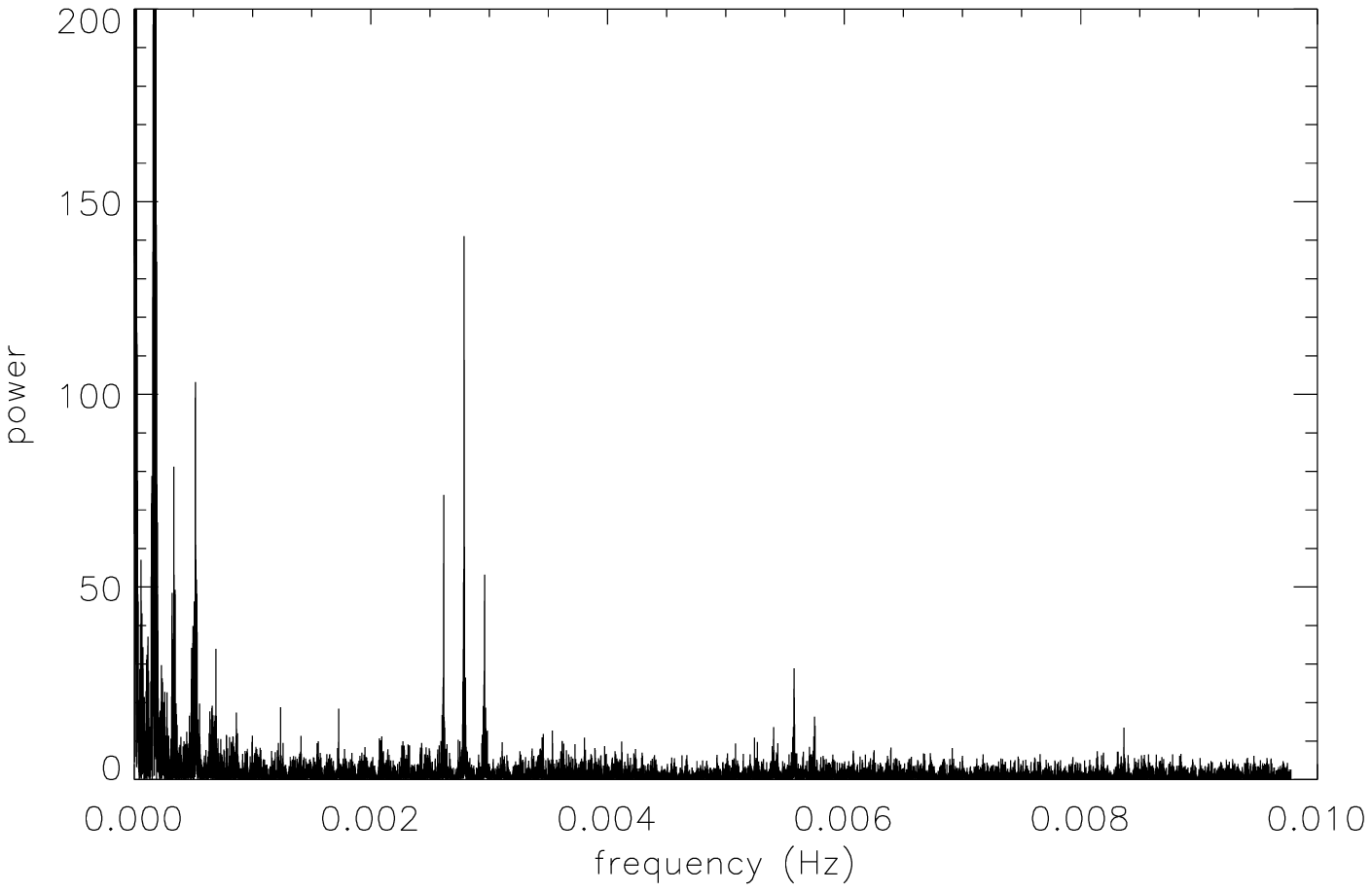,width=\columnwidth,clip=t}

\caption[]{Power spectrum of \bron. 

\label{fig:powspec}}
\end{figure}

A time series was constructed of the intensity of \bron\ in the full
bandpass during MJD 50540 through 50571 with a time resolution of 1 s. 
The times were corrected for the Earth motion in the solar system to 
arrival times at the solar system barycenter. No correction was applied
for the satellite motion around the earth, this results in a maximum error 
of these times of 0.04 s which is not important to the results. A fast Fourier 
transform of the time series resulted in the power spectrum as shown in 
fig.~\ref{fig:powspec}. The spectrum shows an unambiguous peak at 
$2.7883\times10^{-3}$ Hz with side lobes due to the beat with the
Earth occultations each satellite orbit. A second and a third harmonic 
are visible. The peaks to the left are at the (higher harmonics of the)
satellite orbital period.

To check whether the peak in the power spectrum could be an unforeseen 
instrumental effect we calculated the power spectra for the other 3 \mbox{X-ray} 
sources in the field of view (Cyg X-1, X-2 and X-3) as well as for the background.
None of these spectra revealed a peak at the same frequency as observed in the 
power spectrum of \bron.

\begin{figure}[t]
\psfig{figure=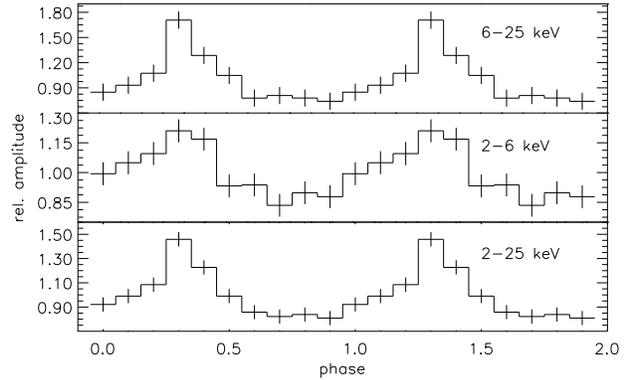,width=\columnwidth,clip=t}

\caption[]{Pulse profile for MJD 50567 through MJD 50571
\label{fig:pulse}}
\end{figure}

The statistical quality of the data does not permit detailed timing analyses,
for instance to search for Doppler delays in pulse arrival times due to
likely binary orbital motion.
We attempted the roughest form of a timing analysis by dividing the best data
in two parts and checking whether a pulse period change could be measured.
The first part applies to all good-quality data between MJD 50543 and MJD 50549. 
This data set encompasses the time of observed peak intensity and is marked 'a'
in fig.~\ref{figlc}. The second set lies between MJD 50567 and MJD 50571, 
and contains the times of a second peak in the time profile. It is marked 'b'
in fig.~\ref{figlc}.  For each time series the period was obtained using an 
epoch folding technique (\cite{Leahy}). Each time series was folded into a 
periodogram with 10 phase bins for 128 trial periods in steps of 0.01 seconds 
around 358.64~s, the peak period from the power spectrum (fig.~\ref{fig:powspec}).
For the first time series the periodogram shows two main peaks at 358.67 and 
358.95 s with smaller peaks to the sides. We therefore estimate the period to be 
$358.81 \pm 0.14$~s. For the second time series the periodogram shows only one 
pronounced peak at 358.61 s. The accuracy of the period of this peak was 
determined through Monte Carlo simulations. The pulse profile was modeled by a 
triangular function with a base width of 30\% duty cycle, in good visual 
agreement with the observed profiles (lower panel in fig.~\ref{fig:pulse}). 
The time series was simulated and a 
periodogram generated 1000 times, each time with a 
different random generator seed value for the time series generation. In each
case the period was determined by that of the highest peak in the 
periodogram. The root mean square of all 1000 periods thus determined prescribes
our estimate of the $1\sigma$ error in the period. It is 0.03~s for the second data set.
The observed change in period is thus $0.20\pm0.14$ s, which is not a 
significant value. We regard the difference between the maximum possible
period in interval 'a' (358.95~s) and the period minus $1\sigma$ of
interval 'b' (358.58) as the upper limit in the period change. It is 0.37~s
over 23 days, or 6~s over 1~yr.

Fig.~\ref{fig:pulse} shows the folded light curve for the second set
'b' in two separate bandpasses, as well as the total bandpass. The amplitude 
is relative to the average intensity. The pulsed
fraction increases with energy, from $\sim 30\%$ at 2--6 keV
to $\sim 50\%$ at 6--25 keV. In the total bandpass it is $\sim40$\%. 
The pulsed fraction is lower in the first set, and averaged over both sets 
it is $\sim30$\% in the total bandpass. We conclude that there is substantial 
variability in the pulsed fraction, although we are not able to measure this 
in detail.

\section{Spectrum}
\label{secspec}

The source is quite weak from the perspective of WFC sensitivity and
the capability to perform a spectral analysis is limited. We fitted
a power law spectrum with absorption by cold interstellar
material of cosmic abundances (according to the model of Morrison \& McCammon 
1983). This was done simultaneously to the data of the seven observations
with the highest signal-to-noise ratio, leaving free per period only the 
normalization. Only data between 2 and 10~keV were used because the detector
response at the remaining energies is not yet fully calibrated. The photon 
index is $1.27\pm0.14$ and $N_{\rm H}=(3.1\pm1.4)\times10^{22}$~cm$^{-2}$ 
(single parameter 1$\sigma$ errors), the reduced
$\chi^2$ is 1.07 for 124 degrees of freedom. A thermal bremsstrahlung spectrum
gives an equally good fit but the temperature is above the WFC bandpass.
Both models imply an unabsorbed peak intensity of 
$8.8\times10^{-10}$~erg~s$^{-1}$cm$^{-2}$ in 2 to 25 keV. The value of $N_{\rm H}$
is about 4 times as high as the Galactic value predicted by interpolation of
maps by Dickey \& Lockman (1990) which suggests circumstellar material in
the line of sight. However, the statistical significance of this
is very limited given the large error on $N_{\rm H}$.

\section{Discussion} \label{sec:discuss}

The transient nature of the source and the photon energies involved strongly 
suggest that \bron\ is an accreting compact object in a binary system. The 
pulsations indicate 
a substantial magnetic field strength and, thus, suggest that the compact 
object is a neutron star or white dwarf. Since there is no firm distance 
estimate for the source (see below), it is not possible to constrain the 
luminosity to be able to discriminate between the neutron star and white dwarf 
nature. Still there are two other characteristics which hint at the nature of
the compact object in \bron. First, although the pulsar signatures of white 
dwarfs in intermediate polars (IPs) are quite similar to those of accreting 
neutron stars in X-ray binaries when one looks at pulse period, pulsed fraction 
and pulse duty cycle (e.g., Norton \& Watson 1989), the pulsed fraction of IPs,
if measured, always increases to lower photon energies. This is opposite to what 
we see in \bron. Second, the Galactic latitude of \bron\ is small which is 
consistent with the distribution of HMXB pulsars along the plane, in contrast
with IPs which are distributed homogeneously across the sky (being relatively 
nearby). If \bron\ belongs to the HMXB pulsar group, its transient nature places 
it in the subgroup of HMXBs with Oe or Be companion stars.

A young object like an early B type star is nearly always found close to its 
birthplace. In 
the line of sight of \bron\ three potential places of recent star formation are
within our galaxy: the OB association Cyg OB7 at approximately
700~pc (Dame \& Thaddeus 1985), the Perseus arm at approximately 4~kpc 
(e.g., Vogt \& Moffat 1975, Georgelin \& Georgelin 1976) and the HI Cygnus arm
at about 11~kpc (Kulkarni, Blitz \& Heiles 1982). A peak flux of 
$8.8\times10^{-10}$~erg~cm$^{-2}$s$^{-1}$ implies respective luminosities of
$5\times10^{34}$, $1.7\times10^{36}$ and $1.3\times10^{37}$~erg~s$^{-1}$ for these three
locations. All of these values are consistent with the Be X-ray binary 
interpretation of \bron.

An optical identification is suggested by the apparent proximity to
HD~200709. For a main sequence B8 star, the absolute
visual magnitude $M_{\rm V}$ is 0.0 (Allen 1973). Regardless of
extinction, an apparent visual magnitude of $m_{\rm V}=9.21$
implies an upper limit to the distance of $7\times10^2$~pc. This distance
appears to be consistent with a membership of HD~200709 to Cyg OB7
although it has not been recognized as such yet. HD~200709 is not
a very good candidate counterpart for two reasons: the positional coincidence
is marginal and it is not recognized as an emission type star as
is common for transient HMXBs, although it should be noted that the
spectral classification appears to be based on multiband photometry only
(Bouigue 1959).

Corbet (1986) discovered a relationship for Be X-ray binaries 
between pulse period and orbital period.
For a pulse period of $358.61$~s this implies an orbital period 
of $\sim$190~d. No evidence was found for a modulation in the
X-ray emission with this period in the WFC data.

The pulse period gives a lower limit to the luminosity through the
propeller effect (e.g., Illarionov \& Sunyaev 1975).
If the Alfv\'{e}n radius is larger than the co-rotation radius (this is
the location where the Keplerian period is equal to that
of the neutron star), infalling matter is prevented from entering the
magnetosphere because of centrifugal forces, and accretion through the
magnetosphere onto the neutron star poles is impossible.
In the case of spherical accretion the lower limit to the luminosity becomes:
$L_{\rm min}=4\times 10^{37} B_{12}^2 P^{-7/3}$ erg s$^{-1}$
(Campana et al. 1998) where $P$ is the spin period in s and $B_{12}$ the neutron star 
magnetic field in units of $10^{12}$ G. Standard neutron star values are
assumed here for radius ($10^6$~cm) and mass (1.4~M$_\odot$).
For a magnetic field of $B=10^{12}$ G and a period of 358.61 s this gives 
$L_{\rm min}=5\times 10^{31}$ erg s$^{-1}$.
This is not very constraining, due to the long pulse period.

To determine whether the measured upper limit on the pulse period
derivative is meaningful, we calculated what maximum derivative is
expected when the accreted matter deposits all of its angular momentum
at the magnetospheric boundary and the magnetic field lines transport
this momentum to the neutron star. Using the expression for the Alfv\'{e}n 
radius by Ghosh \& Lamb (1991) for spherical accretion and assuming standard
values for the neutron star magnetic field, mass,
moment of inertia (10$^{45}$~g~cm$^2$), and radius, the
spin period derivative is given by
${\rm d}P/{\rm d}t=4.1\times10^{-44}L^{6/7}P^2$. For the three
luminosities given above, this derivative indicates a period change
over 23 days of $|\Delta P|=0.006$, 0.1 and 0.8~s. The last possibility
is excluded by our observations which means that either the source is
not at that distance or the change of angular momentum is not as high. We 
conclude that our upper limit for the period derivative is not constraining.

\begin{acknowledgements}
We thank Lucien Kuiper and Frank Verbunt for helpful discussions,
Anton Klumper, Jaap Schuurmans and Gerrit Wiersma for
software support, and the team of the BeppoSAX Science Data
Center in Rome for the major and non-trivial task of generating the many
WFC Final Observation Tapes. BeppoSAX is a joint Italian and Dutch program.
This research has made use of the SIMBAD database operated at CDS, Strasbourg, France.
\end{acknowledgements}

\end{document}